\documentclass[sigplan,screen]{acmart}

\AtBeginDocument{%
  \providecommand\BibTeX{{%
    \normalfont B\kern-0.5em{\scshape i\kern-0.25em b}\kern-0.8em\TeX}}}

\usepackage{graphicx,epstopdf}
\graphicspath{ {./figures/} }
\usepackage{booktabs}
\usepackage{subcaption}
\usepackage{amsmath}
\usepackage{algorithm}
\usepackage{algpseudocode}

\begin{document}
\settopmatter{printacmref=false}
\renewcommand\footnotetextcopyrightpermission[1]{}
\pagestyle{plain}
\title{Global Optimization of Data Pipelines in Heterogeneous Cloud Environments}

\author{Erica Lin*, Luna Xu*, Suraj Bramhavar, Marco Montes de Oca, \\Sean
Gorsky, Lingyun Yi, Arianna Groetsema, Jeffrey Chou}

\affiliation{
	\institution{Sync Computing\\
	\href{info@synccomputing.com}{info@synccomputing.com}}
	\city{Cambridge}
	\state{Massachusetts}
	\postcode{02139}
	\country{USA}
	}
\thanks{Contact author at erica.lin@synccomputing.com}
\newcommand{\proj}{\textsc{AGORA}}

\begin{abstract}
Modern production data processing and machine learning pipelines on the cloud are critical components for many cloud-based companies. These pipelines are typically composed of complex workflows represented by directed acyclic graphs (DAGs). Cloud environments are attractive to these workflows due to the wide range of choice with heterogeneous instances and prices that can provide the flexibility for different cost-performance needs. However, this flexibility also leads to the complexity of selecting the right resource configuration (e.g., instance type, resource demands) for each task in the DAG, while simultaneously scheduling the tasks with the selected resources to reach the optimal end-to-end performance and cost. These two decisions are often codependent resulting in an NP-hard scheduling optimization bottleneck. Existing solutions only focus solely on either problem and ignore the co-effect on the end-to-end optimum. We propose \proj, a scheduler that considers both task-level resource allocation and execution for DAG workflows as a whole in heterogeneous cloud environments. {\proj} first (1) studies the characteristics of the tasks from prior runs and gives predictions on resource configurations, and (2) automatically finds the best configuration with its corresponding schedules for the entire workflow with a cost-performance objective. We evaluate {\proj} in a heterogeneous Amazon Web Services (AWS) cloud environment with multi-tenant workflows served by Airflow and demonstrate a performance improvement up to $45\%$ and cost reduction up to $77\%$ compared to state-of-the-art schedulers. In addition, we apply {\proj} to a real-world production trace from Alibaba and show cost reduction of $65\%$ and DAG completion time reduction of $57\%$.
\end{abstract}

\maketitle
\section{Introduction}
DAG-based data pipelines are commonly used in production environments and have become one of the dominant workloads in the cloud. This is due to their ability to easily describe increasingly complex data analytic workflows caused by increasing business needs. In a DAG, each vertex represents a task and edges encode dependencies and data flow between the tasks. Modern DAGs consist of more than thousands of tasks with complicated dependencies and diverse durations~\cite{2016-graphene,2020-alibaba}. We anticipate both the complexity and the number of DAG workloads to grow exponentially with the inevitable growth of data.

The cloud is widely used to run DAG workloads due to its flexibility, scalability, heterogeneity, and plethora of services. However, executing the same DAG can have drastically different cost and performance characteristics based on resources provisioned, configurations, and scheduling policies even within the same cloud provider. With increasingly complex and large DAGs, this difference can be significant. As a result, there are a lot of works studying VM selection (resource provisioning)~\cite{2017-cherrypick, 2017-vmselecting} and job scheduling~\cite{2014-teris, 2016-graphene, 2020-survay}. However, for a long time this problem has been studied separately with schedulers usually assuming a fixed resource demand from the workloads, in most cases defined by the user. In reality, it is not easy for the user to decide resource demands for each task, and ideally it is a burden that users should not have to carry. Moreover, we found that combining the best solutions in each field does not yield the best solution. This is because these two issues are co-dependent. For example, most VM selection algorithms require runtime estimations which are also heavily dependent on how the jobs are scheduled, while most scheduling algorithms require VM selections to be made prior to scheduling. Optimizing each problem individually does not guarantee a globally optimal solution, and can sometimes even lead to worse results. In this paper, we argue that co-optimization of resource configurations and scheduling is necessary for running complex DAG jobs in heterogeneous cloud environments to achieve the end-to-end optimal goal of cost and performance. 

Co-optimization of resource configurations and job scheduling brings several challenges. Deciding the best workload configuration, VM instances, scale, and schedule is an NP-hard problem. The large set of potential configurations also manifest as an incredibly large search space. In addition, the problem size grows exponentially with the DAG complexity and task diversity. Traditional optimization methods would either require a long time to solve (optimization based approaches) , or can only provide best effort solutions (heuristic based approaches), leaving significant benefit on the table. Recent deep-learning based approaches~\cite{hu_spear_2019} require historical data, training, and model tuning to achieve optimal results. 

In this paper, we propose {\proj}, an automated, globally optimized resource allocator and scheduler for DAG workloads in heterogeneous cloud environments. {\proj} takes one or more DAGs and an optimization goal as input, and co-optimizes the best resource provisions and workload configurations together with task scheduling both within and between DAGs to achieve the best solution. {\proj} supports different goals such as cost, performance, and a balance of both. As opposed to heuristic based approaches, {\proj} is designed to reach an optimal solution, and do so within a reasonable time compared with traditional optimization based approaches. {\proj} first analyzes each task in a DAG and gives a set of runtime predictions associated with different task configurations for big data tasks such as Spark~\cite{2012-spark} configurations. It then formulates the scheduling problem as a variation of the resource-constrained project scheduling problem (RCPSP~\cite{herroelen_project_2009}) in which the task demands and runtimes are also expressed as variables for optimization. Finally, it solves the optimization problem and outputs an optimal resource configuration for each task and an optimal schedule for all the DAGs. 

Specifically, the paper makes the following contributions:
\begin{enumerate}
    \item We present a motivational study to show that existing separate optimization does not lead to the best global optimum. We also quantify the challenge of co-optimization.
    \item We formally define the co-optimization problem and extend existing formulation to include both resource configuration and job scheduling.
    \item Based on the problem formulation we design a solver that achieves low overheads, and implement {\proj} to demonstrate our algorithm.
    \item We evaluate {\proj} with micro and macro benchmarks and show that {\proj} is able to both improve job makespan up to $45\%$ and cost up to $77\%$. 
\end{enumerate}
\section{Related Work}
Running data pipelines in the cloud currently involves two steps: resource allocation (selecting the right VM instance types and deciding the cluster size) and job scheduling. Both have been long-lasting topics in research~\cite{2020-survay}. However, the synergy and codependency between these two have not been studied intensively.

\subsection{Resource Allocation}
\label{sec:related-resource}
The cloud offers heterogeneous compute resources at various prices. It is important to decide the right resource configurations for each task to achieve the best cost-performance for a given goal. Existing methods apply runtime prediction to predict the runtime and calculate the cost with different instances. With the calculated runtime and cost, the resource allocator is able to pick the right types and number of instances which come closest to the goal (e.g., lowest cost, lowest runtime, etc.). Here we provide an overview of some recent predictors. Ernest~\cite{2016-ernest} is a predictor that can estimate runtime of the same job executed on a range of different numbers of machines, reporting a low average prediction error of less than $20\%$. It utilizes a general model and a nonlinear least squares (NNLS) solver to fit the runtime data. Cherrypick~\cite{2017-cherrypick} is an example of a predictor that adopts a black box approach as opposed to Ernest. Cherrypick requires multiple runs of the same application in order to do the prediction. Some predictors utilize an analytical model approach that can make predictions on data from one run of the application. Wang et al.~\cite{wang_performance_2015} present an implementation using a model that predicts a Spark application's runtime on a stage by stage basis by taking into account the stage overheads, task overheads, and task runtimes. While this approach has fewer data collection requirements, the accuracy is lower than that of Ernest or Cherrypick. Ardagna and Pinto~\cite{ardagna_performance_2018} utilize a simulator approach to predict application runtimes, achieving low prediction error, however it does not allow for the potential of accounting for parameters in big data frameworks such as Spark. There are also tools focused on autotuning Spark parameters, but they do not predict performance for different hardware~\cite{buchaca_you_2020}. Finally, Sayeh et al.~\cite{al-sayeh_gray-box_2020} take a gray-box modeling approach, using both a white-box model and a black-box model. They achieve impressive results in prediction accuracy, but their model requires running the application many times with many different configurations.

\subsection{Job Scheduling}
Given resource demands and the runtime of each task, the scheduler then orders and packs the tasks on the underlying cluster for execution. This paper summarizes recent scheduling algorithms into two categories: heuristic based approaches and optimization based approaches. 

Heuristic based methods (e.g. FIFO, round-robin, shortest job first, critical path, etc.) have been long adopted in operating systems for job scheduling, and are also widely applied in distributed systems. Tetris~\cite{2014-teris} made an early attempt to pack multi-resource tasks to a cluster of resources for maximizing utilization. It adopts a shortest-remaining-job-time-first heuristic to improve performance. Tetris achieves both high cluster utilization and performance, but it is not DAG-aware. Graphene~\cite{2016-graphene} extends Tetris and embeds task dependency into the scheduling. Graphene schedules long-running or hard-to-pack tasks first. Although the heuristic seems simple, Graphene requires multi-layer nested iterations to optimize for the best schedules which requires long solving times. Carbyne~\cite{2016-altruistic} is another DAG-aware fair scheduler that utilizes leftover resources to further reduce resource fragmentation. All these schedulers assume predefined resource usage for each task, and they only work with homogeneous cloud environments (i.e., they are not aware of the different instance types, and they do not select the right VMs for the tasks). Stratus~\cite{2018-stratus} on the other hand is the closest work to {\proj} in the sense that it selects VMs for each task to minimize cost. In addition, it schedules workloads with similar runtimes to maximize VM utilization. Although Stratus selects and launches VMs for a task based on the runtime estimation, it still assumes a predefined resource demand {\it(vcpu, memory)} whereas {\proj} decides the configurations including VM instance types and resource demands for each task together with the best schedule. Moreover, Stratus is not DAG-aware and it only optimizes cost while {\proj} can optimize for different objectives. Apart from the aforementioned offline schedulers that use workload runtime estimation or prior knowledge to achieve near-optimal results, online schedulers~\cite{2018-kairos, 2018-rupam, yabuuchi2020multi, agrawal2016scheduling} require no prior task runtime information and adaptively schedule tasks during runtime. These approaches are orthogonal and could be applied as complementary to {\proj}.
\begin{figure}[t]
\centering
\includegraphics[width=0.9\linewidth]{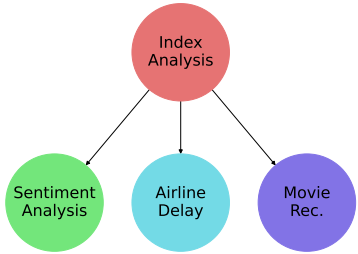}
\vspace{-20pt}
\caption{Example DAG composed by real-world data analytic pipeline jobs.}
\label{fig:motivation1_dag}
\end{figure}

Optimization based methods formulate cluster scheduling into an optimization problem, and strive to solve the problem to achieve the optimal solution. These methods could differ in both the problem formulation and the solver. Most commonly, the optimization problem is formulated as a general Linear Programming (LP~\cite{chen_minimum_2019, bodik_brief_nodate}) problem, Mixed Integer Linear Programming (MILP~\cite{1997-milp})
problem, or a flow-based graph optimization problem such as Min-Cost Max-Flow (MCMF~\cite{ahuja_magnanti_orlin_1993}) problem. For example, TetriSched~\cite{2016-tetrisched} automatically translates the workload resource requests as a MILP problem and solves it to effectively schedule tasks. Firmament~\cite{gog_firmament_nodate} and Quincy~\cite{isard_quincy_2009} both formulate the problem as an MCMF problem and focus on fairness, data locality, and scheduling latency. Similar to the heuristic based schedulers, these works are not DAG-aware, and they assume predefined resource demands instead of co-optimizing. There are also schedulers that consider task dependencies and allow for resource demands to be malleable~\cite{bodik_brief_nodate, chen_minimum_2019}, but they do not consider cost in the optimization and are not heterogeneity-aware. Also, for simplicity they neglect workload characteristics and assume all tasks will run faster with more resources. As a special case of optimization based approaches, deep learning and machine learning based approaches~\cite{hu_spear_2019,mao_resource_2016} have been shown to find good solutions if trained with adequately sized data sets, but even these approaches often result in solutions which are sub-optimal and may shift over time as real workloads deviate from the training data sets.  Similarly, if provisioning and scheduling are performed as two independent functions, even the perceived "optimal" solution calculated by any advanced scheduling computation will not, in fact, represent the best possible solution for the entire system.  {\proj} aims to overcome this limit by combining both functions into one calculation.
\section{Motivation}
\label{sec:motivation}

\begin{table}[t]
\caption{Selected instance types from AWS. The spec and price information are valid on 01/27/2022.}
\vspace{-8pt}
\label{tab:instance}
\begin{tabular}{@{}llll@{}}
\toprule
Instance    & vCPUs & Memory & Cost (\$ per hour) \\ \midrule
m5.4xlarge  & 16    &  64      &0.768      \\
m5.8xlarge  & 32    & 128       & 1.536     \\
m5.12xlarge & 48    & 192       & 2.304     \\
m5.16xlarge & 64    & 256       & 3.072    \\ \bottomrule
\end{tabular}
\vspace{-10pt}
\end{table}

\begin{figure*}[h]
    \centering
    \includegraphics[width=\linewidth]{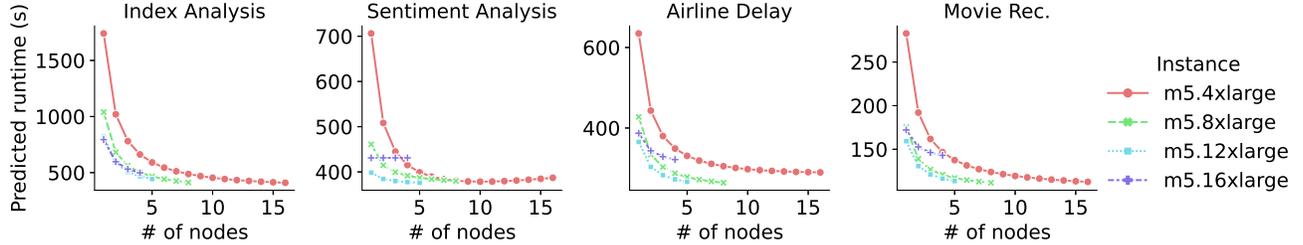}
    \vspace{-30pt}
    \caption{Ernest runtime prediction on four example jobs with selected instance types. X-axis shows the number of nodes used for each instances. Y-axis shows the predicted runtime from Earnest.}
    \label{fig:ernest}
\end{figure*}

In this section, we first demonstrate the need for co-optimizing resource allocation and task scheduling with a simple DAG example. We also showcase the challenges of co-optimization with increasing DAG complexity. For the first demonstration, we select four applications in real-world data production pipelines: Index Analysis, Sentiment Analysis, Airline Delay, and Movie Recommendation. Index Analysis is a common data pre-processing job that reads raw data from storage, extracts features, and writes back to S3 for analytic jobs further down the pipeline. The other three jobs are typical ML jobs that analyze data for predicting airline delays, perform text sentiment analysis with NLP, and recommend movies, respectively. For simplicity, we form a DAG as shown in Figure~\ref{fig:motivation1_dag} as our example. The DAG demonstrates a typical data analytic pipeline (i.e., three ML jobs after data pre-processing). We select four instance types from AWS cloud~\cite{aws} as shown in Table~\ref{tab:instance}. In reality, DAGs are more complicated with more variable instances types, which makes the problem more difficult and results in worse performance than our example. 

\begin{figure*}
    \begin{subfigure}[t]{0.34\textwidth}
    \centering
    \includegraphics[width=1.52\linewidth]{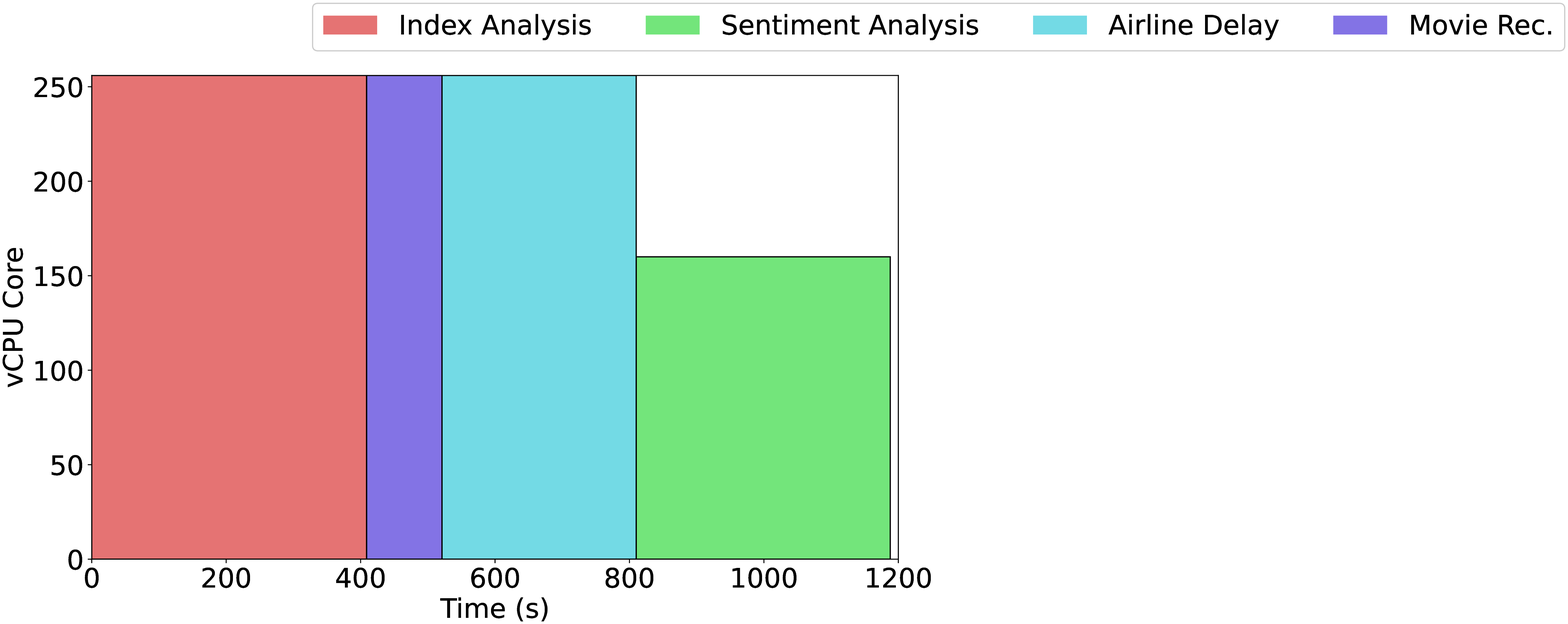}
    \caption{{\it separate} optimization approach.}
    \label{fig:motivation1_sota}
    \end{subfigure}   
    \begin{subfigure}[t]{0.31\textwidth}
    \centering
    \includegraphics[width=\linewidth]{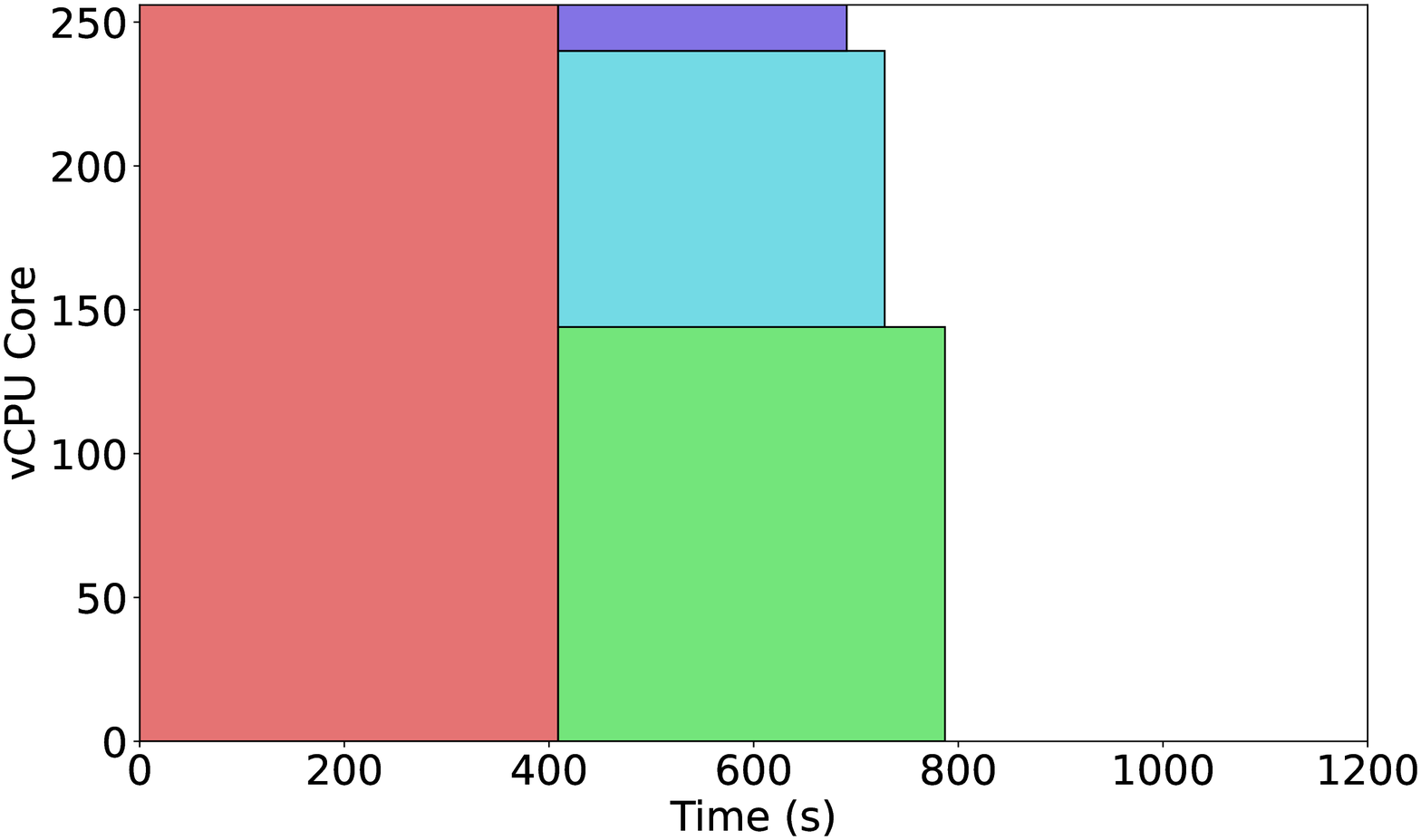}
    \caption{{\it Brute-force (BF) co-optimize} approach.}
    \label{fig:motivation1_oracle}
    \end{subfigure}    
    \begin{subfigure}[t]{0.32\textwidth}
    \centering
    \includegraphics[width=\linewidth]{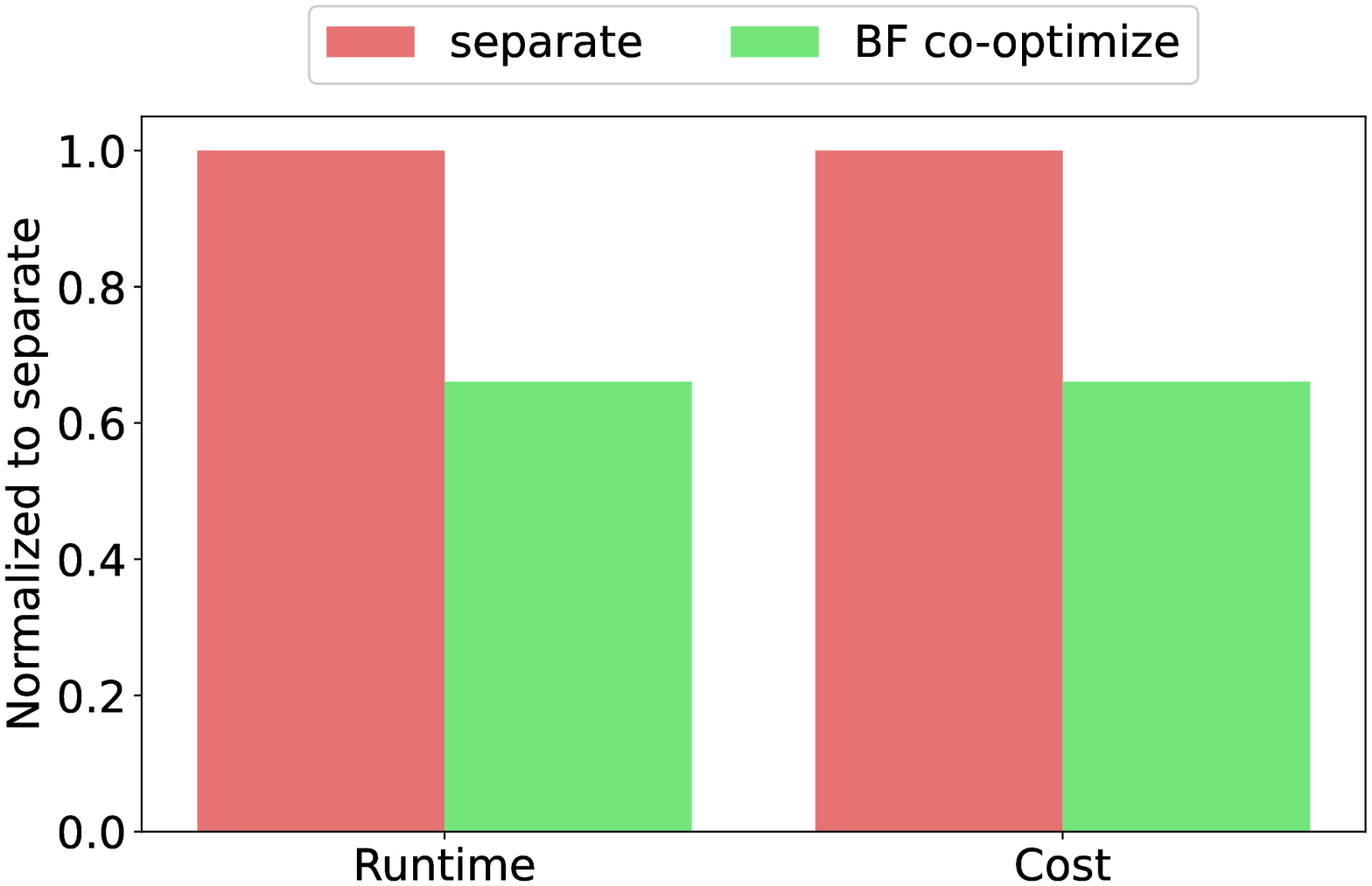}
    \caption{Overall normalized DAG runtime and cost.}
    \label{fig:motivation1_improvement}
    \end{subfigure}     
    \vspace{-5pt}
    \caption{Runtime of {\it separate} and {\it co-optimize} with job scheduling breakdown and the overall DAG runtime and cost.}
\end{figure*}

\textbf{Best of each world does not bring the best outcome:} One intuitive approach to run a DAG in the cloud is to first decide resource configurations for each task and select the best VM instances to run the DAG. To simulate the best effort in this step, we selected Ernest~\cite{2016-ernest}, a performance predictor for big data jobs, to help us decide the best configurations. Ernest is reported to achieve less than $20\%$ error on most workloads with less than a $5\%$ overhead for training, making it a highly efficient and accurate predictor. We run Ernest with our example DAG and Figure~\ref{fig:ernest} shows the results. We evaluate with the goal of shortest runtime in this set of experiments. From Figure~\ref{fig:ernest}, we see mostly diminishing returns in runtimes with increasing number of nodes for every instance type, with Sentiment Analysis even showing negative scaling on a large number of m5.4xlarge instances. We reasonably select VM configurations according to our goal of shortest runtime as shown in Table~\ref{tab:vm-select}. After selecting the VMs, the DAG is usually submitted to a scheduler for execution. The scheduler typically reads DAG dependencies and resource configurations to decide job packing and execution order. A good scheduler could further reduce the overall DAG runtime, which results in lower cost. In this experiment, we select TetriSched~\cite{2016-tetrisched} as our scheduler with modifications to allow for dependency-aware scheduling. TetriSched is an optimization-based scheduler, using a Mixed Integer Linear Programming (MILP) formulation, that can solve to a proven optimal solution. Combining the results of Ernest and TetriSched, we simulate a separately optimized approach ({\it separate}) for running a DAG in cloud. The result of this approach is shown in Figure~\ref{fig:motivation1_sota}. 
\begin{table}[t]
\caption{VM selection configurations for all jobs under {\it Ernest} and {\it BF co-optimize}.}
\vspace{-8pt}
\label{tab:vm-select}
\begin{tabular}{@{}lll@{}}
\toprule
Jobs               & \textit{Ernest} & \textit{BF co-optimize} \\ \midrule
Index Analysis     & 16 x m5.4xlarge  &  16 x m5.4xlarge  \\
Sentiment Analysis & 10 x m5.4xlarge  & 9 x m5.4xlarge       \\
Airline Delay      & 16 x m5.4xlarge  & 6 x m5.4xlarge       \\
Movie Rec.         & 16 x m5.4xlarge  & 1 x m5.4xlarge       \\ \bottomrule
\end{tabular}
\vspace{-15pt}
\end{table}

Although the {\it separate} approach optimizes each step carefully, it does not result in the overall best outcome. To demonstrate this, we perform an exhaustive search on all possible combinations of VM configurations with DAG scheduling together for a global minimum runtime (referred to as brute-force co-optimization {\it BF co-optimize}). Table~\ref{tab:vm-select} shows the resulting VM configurations from the exhaustive search and Figure~\ref{fig:motivation1_oracle} shows the runtime. We can see that {\it BF co-optimize} reaches a $40\%$ improvement in runtime and cost (Figure~\ref{fig:motivation1_improvement}). This is because {\it BF co-optimize} combines scheduling with VM selection to reach a global optimal. From Figure~\ref{fig:motivation1_sota} and Figure~\ref{fig:motivation1_oracle} we see that though individually the three ML jobs are running longer than {\it separate} with reduced resource configurations, the scheduler is able to overlap them to achieve a global optimal DAG runtime. 

\begin{figure}
    \centering
    \includegraphics[width=8.5cm]{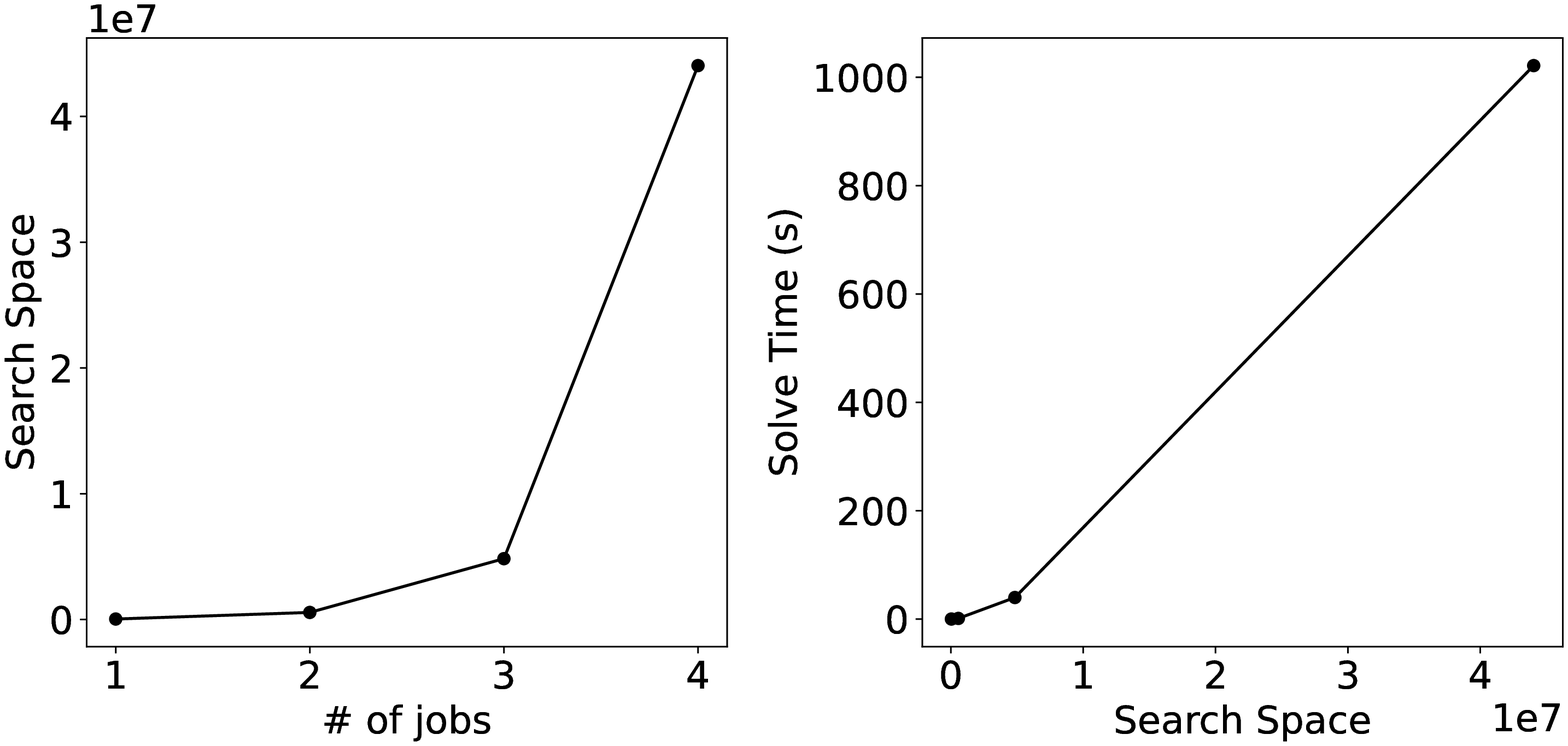}
    \caption{Search space and solve time grows exponentially with the number of jobs grows in a DAG for {\it BF co-optimize}. Here the search space is measured by how many values to search from.}
    \label{fig:motivation2}
    \vspace{-15pt}
\end{figure}
\textbf{Co-optimization performance bottleneck:} Previous experiments show the necessity of co-optimizing resources and scheduling when running DAG workflows for best cost and performance. However, co-optimization is a non-trivial NP-hard problem. We are not aware of any existing solution that considers such a co-optimization. Although the brute-force approach via exhaustive search is commonly performed to find the best solution~\cite{2016-graphene}, it suffers from a heavy performance bottleneck that can diminish the benefits. In this experiment, we demonstrate the challenge of co-optimization with exhaustive search when the problem size increases. For simplicity, we only explore the complexity of a single DAG with increasing number of jobs. In reality, the relationship between jobs and number of VM instance types could further grow the problem size significantly. The left graph of Figure~\ref{fig:motivation2} shows that when number of jobs in a DAG increases, the problem size (i.e., search space) grows exponentially. Only four jobs in a DAG could result in tens of millions of values to search from, and this is ignoring the factor of VM instance types. The right graph of Figure~\ref{fig:motivation2} shows how long it takes to solve a problem with existing search methods when problem size grows. Note that this is only a simple example that ignores VM instance types.

In conclusion, we use a simple example to show how separately optimizing resource configurations and job scheduling does not necessarily bring the best outcome. Simultaneously, co-optimization complicates the problem and can lead to a significant overheads which could undermine its benefit. There is a need for a holistic solution that can yield a globally optimal cost-performance while maintaining reasonable overhead requirements.

\section{{\proj}}
{\proj} is an automated globally optimized resource allocator and scheduler designed to optimize performance and cost for DAG workloads in the cloud. {\proj} co-optimizes resource configurations and job scheduling to reach a global optimum for DAG workloads with configurable objectives (e.g., lowest cost, shortest runtime, etc.). {\proj} is carefully designed to address the performance bottleneck of the brute-force approach and aims to achieve the global optimum within negligible optimization overheads. We design {\proj} and demonstrate the idea of co-optimization with DAGs composed of Spark~\cite{2012-spark} jobs due to its popularity and impact. However, the idea of co-optimization is not limited to Spark jobs and the algorithms described can be applied to other DAG-based jobs.
In this section, we first describe the overall architecture of {\proj}, followed by detailed description of the algorithm {\proj} adopts to co-optimize the resource configuration and job scheduling.

\subsection{System Architecture}
\begin{figure*}
    \centering
    \includegraphics[width=0.75\linewidth]{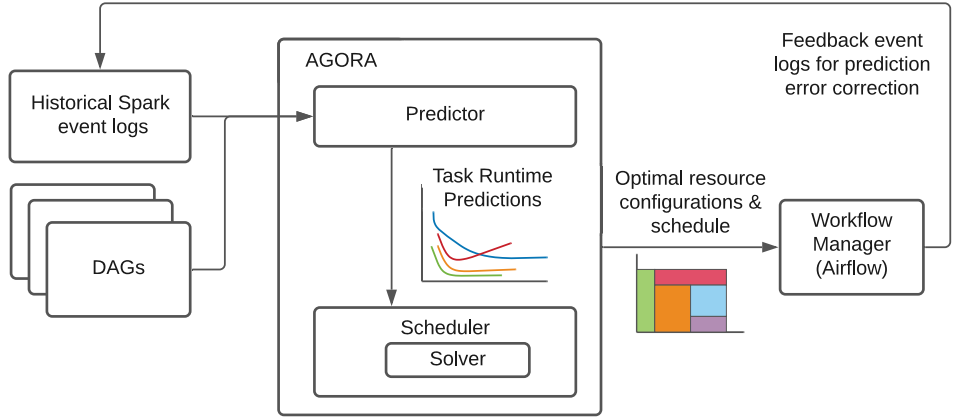}
    \caption{System architecture and workflow of {\proj}.}
    \label{fig:agora}
\end{figure*}

Figure~\ref{fig:agora} shows an overall system architecture and workflow of {\proj}. {\proj} has two key components: {\bf Predictor} and {\bf Scheduler}. Similar to the aforementioned two-step approach to run DAGs in cloud, the {\bf Predictor} component is in charge of job runtime prediction with different resource configurations, and the {\bf Scheduler} component decides job scheduling. However, different from the two-step approach, {\proj} optimization algorithm always involves both components instead of optimizing separately.

After users submit DAGs for execution, {\proj} reads the DAGs and identifies the tasks in each DAG. The {\bf Predictor} reads a historical event log provided by Spark for each task, and predicts the runtime of a job under different configurations. Note that different from other runtime predictors described in Section~\ref{sec:related-resource}, {\proj} requires only one event log per task (one prior run). This historical event log can be provided by users or gathered by {\proj} with a triggered test run. {\proj} saves the event log into a database for future reference. With the runtime predictions, {\proj} continually optimizes with the {\bf Scheduler} to reach a globally optimal resource configuration and job schedule for the entire DAG(s) execution. Here, the schedule and task configurations are represented as a single problem to be co-optimized. The solver within the {\bf Scheduler} component solves the co-optimization problem. Note that {\proj} supports optimization for one DAG as well as multiple DAGs, resulting in the best performance for all DAGs. Users are able to configure {\proj} with an optimization goal to fit different needs. The resulting resource configurations and schedules are then passed down to a workflow manager (e.g., Airflow~\cite{airflow}) for job execution. After execution of the DAGs, new event logs are collected and fed back to the {\bf Predictor} to improve the accuracy of the prediction model. In production pipelines where the same DAG is periodically executed (e.g. once a day), {\proj} can continue to improve itself with better runtime predictions and converge to better solutions. 


\subsection{Problem Formulation}
\label{sec:design-problem}
\begin{table}[]
\caption{Glossary of notations.}
\vspace{-8pt}
\label{tab:variables}
\begin{tabular}{@{}ll@{}}
\toprule
Notation & Description \\\midrule
$M$ & original makespan       \\
$M_{opt}$ & optimal makespan \\
$M_{budget}$ & makespan budget \\
$C$ & original cost \\
$C_{opt}$ & optimal cost \\
$C_{budget}$ & cost budget \\
$w$ & makespan weight \\
$ts_{ij}$ & start time of task j in DAG i \\
$te_{ij}$ & end time of task j in DAG i \\
$d_{ijc}$ & runtime of task j in DAG i for configuration c \\
$r_{jtmc}$ & resource demands of task j at time t for resource m \\
& for configuration c \\
$R_{m}$ & total capacity of resource m \\
$C_{m}$ & cost of resource m \\
$D$ & set of all DAGs \\
$T_i$ & set of all tasks in DAG i \\
$P$ & set of all precedence constraints \\
$N$ & set of all resources \\
$F$ &flip probabilities\\
\bottomrule
\end{tabular}
\vspace{-20pt}
\end{table}
This section formally describes the co-optimization problem {\proj} solves. Table~\ref{tab:variables} lists the notations used in this section. {\proj} is designed to address the following problems: 
Given a set of DAGs $D$ to be run in a set of heterogeneous resources $N$ and an optimization goal, (1) what is the resource configuration (VM instance, number of VMs, application specific parameters) for each task in the DAG, and (2) what is the order to launch tasks in the DAGs to achieve the global end-to-end optimization goal. Other than VM selection, {\proj} also considers application specific parameters such as Spark configurations because we found that these configurations can directly decide the resource usage per task (e.g. executor memory) and have a big impact on the runtime. 

Formally, the problem can be formulated as a resource-constrained project scheduling problem (RCPSP~\cite{herroelen_project_2009}). A RCPSP consists of scheduling tasks with given resource demands, durations, and precedence constraints on a set of resources with given capacity limits. The RCPSP formulation accurately captures the flexibility of constraints as well as the integrated nature of the assignment and scheduling problem. For example, tasks are allowed to execute on any number of resources as opposed to a specific machine in job shop scheduling~\cite{manne_job-shop_1960}. The RCPSP representation also allows us to not have to decompose the problem into two sub-problems (e.g. assign, then sequence) as common in other formulations~\cite{brandimarte_routing_1993}. Therefore, we do not have to make decisions about which machine each task has to run on prematurely prior to optimization, resulting in better solutions.  As opposed to classic formulations of RCPSP where demands and runtimes for each task are given and fixed, {\proj} requires the resource demands and task runtimes to be variables to allow the solver to optimize the resource configurations and schedules jointly. To this end, we extend the original RCPSP problem to suit the need of {\proj}. More specifically, we formulate the RCPSP problem as follows:
\begin{equation} 
\text{minimize } \frac{M_{opt}-M}{M}*w + \frac{C_{opt}-C}{C}*(1-w) \text{ subject to } \label{eq:objective}
\end{equation}

\begin{equation}
te_{ij} = ts_{ij} + d_{ijc}, \forall j\in T_i, \forall i\in D \label{eq:process}
\end{equation}

\begin{equation}
ts_{ij} \geq te_{ik}, \forall i\in D, \forall (j,i)\in P, \label{eq:precedence}
\end{equation}

\begin{equation}
\sum_{t=0, \forall j\in T_i}^{M} r_{jtmc} \leq R_m, \forall m \in N, \forall i\in D,  \label{eq:resource}
\end{equation}

\begin{equation}
te_{ij} \leq M_{opt}, \forall j\in T_i, \forall i\in D, \label{eq:makespan}
\end{equation}

\begin{equation}
C_{opt} = \sum_{\forall j\in T_i, \forall i\in D, \forall m\in N} r_{jtmc}*d_{ij}*C_{m}, \label{eq:cost}
\end{equation}

\begin{equation}
M_{opt} \leq M_{budget}, \label{eq:makespan_budget}
\end{equation}

\begin{equation}
\text{and } C_{opt} \leq C_{budget}, \label{eq:cost_budget}
\end{equation}

Equation~\ref{eq:objective} defines the optimization objective, where the improvement in makespan and cost are minimized according to a tunable weight parameter $w$. The weight parameter indicates how important the makespan is relative to cost. For example, the user could set $w=0.5$ to achieve a balanced optimization between runtime and cost; $w=0$ to optimize for the lowest cost and $w=1$ to achieve the shortest runtime. Users can configure $w$ between $0$ and $1$ to allow for a flexible optimization preference. Equation~\ref{eq:process} defines the task start and end times with the predicted runtime with different configurations. We have relaxed the classic RCPSP to allow for the runtime to be malleable. This is the key change of the RCPSP formulation to allow {\proj} for co-optimization of resource configurations and scheduling.  
Equation \ref{eq:precedence} defines the precedence constraints given by the DAGs where a task can only start when the precedent task finishes, allowing {\proj} to be DAG-aware. Equation~\ref{eq:resource} defines the resource constraints to ensure that the sum of demands of scheduled tasks do not exceed the capacity of the cluster. Here, a resource can be any cluster capacity constraint, including number of cores, memory, or bandwidth. The optimal makespan is defined by Equation \ref{eq:makespan}, where the end time of each task is less than the optimal makespan. Finally, Equations~\ref{eq:makespan_budget} and~\ref{eq:cost_budget} ensure that the makespan and/or cost are within user-defined budgets. The makespan and cost budget can be optimally set by the user. If either are not set, they are set with a default value of infinity.

In our problem formulation, we adopt a simplified cost model where the cost is determined by a product of the resources demanded, runtime, and cost of the instance type. This assumes that other cloud resources, such as storage, remain the same for all configurations. In reality, costs in the cloud are increasingly complex, and the cost equation can be replaced by more representative cost models depending on the user. For example, spot instances in AWS have a dynamic pricing model that fluctuates based on the current market's demands. {\proj} can be easily modified to include these details by defining the $C_m$ variable more accurately. 

\begin{algorithm}[t]
\caption{Resource allocation and scheduling co-optimization algorithm}\label{alg:opt}
\begin{algorithmic}
\While{not stopping criterion}
\State $c \gets get\_new\_configuration(c)$
\State $M_{new}, C_{new} \gets SAT\_Solver(c, d, P, R)$ \Comment{optimal}

\State $\Delta E \gets calculate\_delta\_energy(M_{new}, C_{new})$
\If{$\Delta E < 0$}
    \State $F \gets 1$
\Else
    \State $F \gets exp(-\Delta E / T)$
\EndIf

\If{$F > acceptance probability$}
    \State $save\_configurations()$
    \State $M_{opt} = M_{new}$
    \State $C_{opt} = C_{new}$
\EndIf
\EndWhile
\end{algorithmic}
\end{algorithm}

\subsection{Optimization Solver}
Classic RCPSP problems are NP-hard~\cite{blazewicz_scheduling_1983}. To apply the RCPSP formulation to our problem, we add resource configurations and job runtimes as additional variables, which makes the problem harder to solve and can lead to long solve times. To address this challenge, {\proj} utilizes a combination of a simulated annealing algorithm~\cite{kirkpatrick_SA} and a satisfiability (SAT~\cite{horbach_boolean_2010}) solver to find optimal solutions in a reasonable amount of time. In this implementation, we use CP-SAT from Google OR-Tools~\cite{ortools} as the SAT solver component. The CP-SAT solver combines a finite domain propagation engine typical for constraint programming (CP~\cite{rossi2006handbook}) solvers with the advantages of advanced SAT solvers by utilizing lazy clause generation~\cite{hutchison_lazy_2010}. Lazy clause generation allows the solver to lazily create the SAT clausal representation of the problem as the computation progresses, therefore reducing the total number of clauses and variables required at any given time. The CP-SAT solver has been shown to outperform other solvers on classic RCPSP benchmarks~\cite{schutt_explaining_2011}. 

Simulated annealing is often used in bound-constrained optimization problems~\cite{rios_derivative-free_2013}. The algorithm mimics the physical process of heating material then slowly lowering the temperature to decrease defects, resulting in a minimum energy state. Similar to other greedy algorithms, a new point is accepted if it results in a lower objective, but it also avoids getting trapped in local minimum by accepting points that result in a higher objective at a decreasing probability. In {\proj}, simulated annealing is used to optimize for the resource configurations given the results of the schedule optimization, while the SAT solver optimizes the schedule given specific resource configurations for each application. Algorithm~\ref{alg:opt} shows the detailed steps of {\proj}. In each iteration, a new set of resource configurations $c$ is chosen for a set of tasks. The solver then minimizes the objective for the current set of resource configurations with a corresponding new optimal makespan $M_{new}$ and cost $C_{new}$. Here, the optimization problem is solved to the global optimal solution. The new set of configurations and schedules are accepted based on acceptance probabilities and flip probabilities ($F$) dependent on the energy difference ($\Delta E$) between the current best solution and the new solution. If the energy difference is negative (e.g. new solution has a better objective than the current best solution), then the new solution is accepted. Otherwise, the new solution is accepted if the probability is greater than a randomly generated acceptance probability, allowing the search to escape from local minima. The search continues until a stopping criterion (e.g. time limit or convergence). This allows us to stop the search when there are diminishing returns. Each component of the solver depends on the other to reach an optimal solution, and together, they jointly optimize for both the schedule and resource configurations. The time complexity of simulated annealing algorithms heavily depends on the cooling schedule, which determines the number of iterations required. Here, because we define the objective as a sum of the percentage of improvements in cost and runtime, we can define a constant starting annealing temperature of $1$ for all problem sizes. The cooling rate we define as a function of $n$, and we define a fixed convergence criteria, resulting in a time complexity of the simulated annealing algorithm of $O(n)$. Within each iteration of the simulated annealing iteration, we utilize the SAT solver. While the worst case time complexity of SAT solvers are exponential, they can efficiently solve problems with tens of millions of variables and constraints with techniques like conflict-driven clause learning (CDCL~\cite{marques2021conflict}) . In addition, lazy clause generation helps keep SAT problem sizes small. By carefully selecting and designing the solver, {\proj} can reach a optimal solution within negligible overheads. Section~\ref{sec:eval-overhead} has a detailed overhead and scalability analysis of {\proj}.

\subsection{Predictor}
As described before, runtime prediction is a key component to solving the problem. {\proj} does not limit the choice of runtime predictor. Users can plugin any predictor mentioned in Section~\ref{sec:related-resource}. In this paper, we design an in-house {\bf Predictor} for Spark jobs that not only selects VMs, but also decides Spark specific parameters to accurately fine-control the resource usage for each task. The {\bf Predictor} takes in a historical Spark event log from a previous run and predicts the application's runtime for a set of infrastructure hardware (i.e., instance types and number of instances) and Spark configurations (i.e., number of executors, executors per core, and memory per core). The {\bf Predictor} utilizes both an analytic model and simulation to forecast the runtime by predicting changes in task runtime distributions and simulating how the application will run on different sets of infrastructure and Spark configurations. Details on the {\bf Predictor} is out of the scope for this paper. In general, {\proj} is heavily dependent on the accuracy of runtime prediction. To provide better results, {\proj} adopts an adaptive approach by collecting new event logs and adjust the model to improve the prediction error.

\section{Evaluation}
\begin{figure}[t]
\centering
\includegraphics[width=\linewidth]{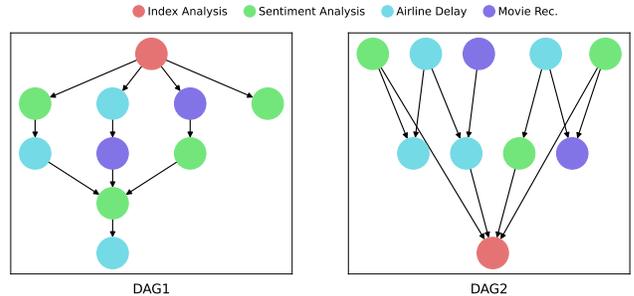}
\caption{Selected DAGs for evaluation.}
\label{fig:eval_dag}
\vspace{-15pt}
\end{figure}

In this section, we evaluate {\proj} on a heterogeneous AWS cloud environment (Table~\ref{tab:instance}). We have implemented {\proj} in Airflow~\cite{airflow}, a widely adopted platform for running DAG workflows in distributed systems. Airflow internally calculates job priority weights by how many children a job has in a DAG and schedules jobs accordingly. FIFO heuristic is applied when multiple jobs have the same topological order (i.e., same priority weights). We embedded {\proj} into Airflow ({\it ver. 2.2}) by adding a configurable {\proj} scheduling policy. We also modified the Spark operator in Airflow to support {\proj} Spark configurations. We select two representative DAGs as shown in Figure~\ref{fig:eval_dag} to demonstrate the effect of {\proj} on different DAG shapes. DAG1 mimics a common workflow where data is pre-processed, then utilized by different machine learning workloads that build on each other. There are several bottlenecks in the DAG that could occur when a single task depends on multiple different tasks to combine all the results into something useful to the organization. DAG2 on the other hand, mimics a workflow where machine learning workloads that build on each other are performed first and converge in a final data analysis application. In this case, many tasks can run in parallel and the only bottleneck is the final task. Both DAGs are composed with real-world Spark jobs described in Section~\ref{sec:motivation}. We carefully choose the Spark configurations for each job to achieve best performance with careful experiments and recommendations from Spark experts. Note that {\proj} tunes Spark configurations based on the characteristics from historical log. The experiments are run with Airflow {\it ver. 2.2.3} and Spark {\it ver. 3.2.0}, which are the latest versions at the time of writing this paper. We evaluate both the end-to-end runtime and cost of each DAG to demonstrate the efficacy of {\proj}. Finally, we also apply {\proj} on a real-world large-scale trace from Alibaba~\cite{2020-alibaba} to evaluate how {\proj} would perform in a multi-DAG production environment.

\subsection{Overall Performance}
\label{sec:overall}
\begin{figure*}[t!]
    \begin{subfigure}[t]{0.32\linewidth}
    \centering
    \includegraphics[width=\linewidth]{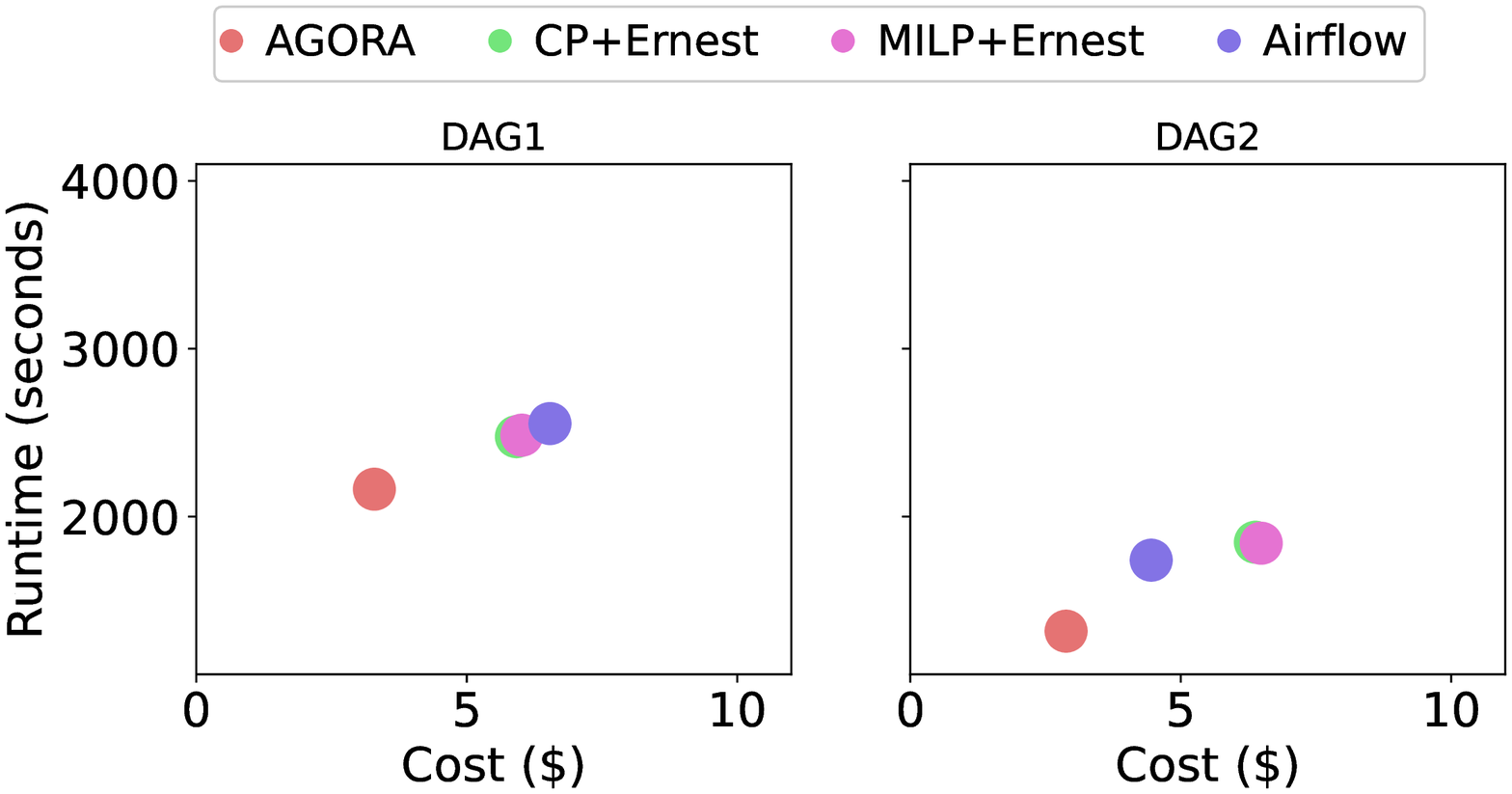}
    \caption{balanced}
    \end{subfigure}   
   \begin{subfigure}[t]{0.32\linewidth}
    \centering
    \includegraphics[width=\linewidth]{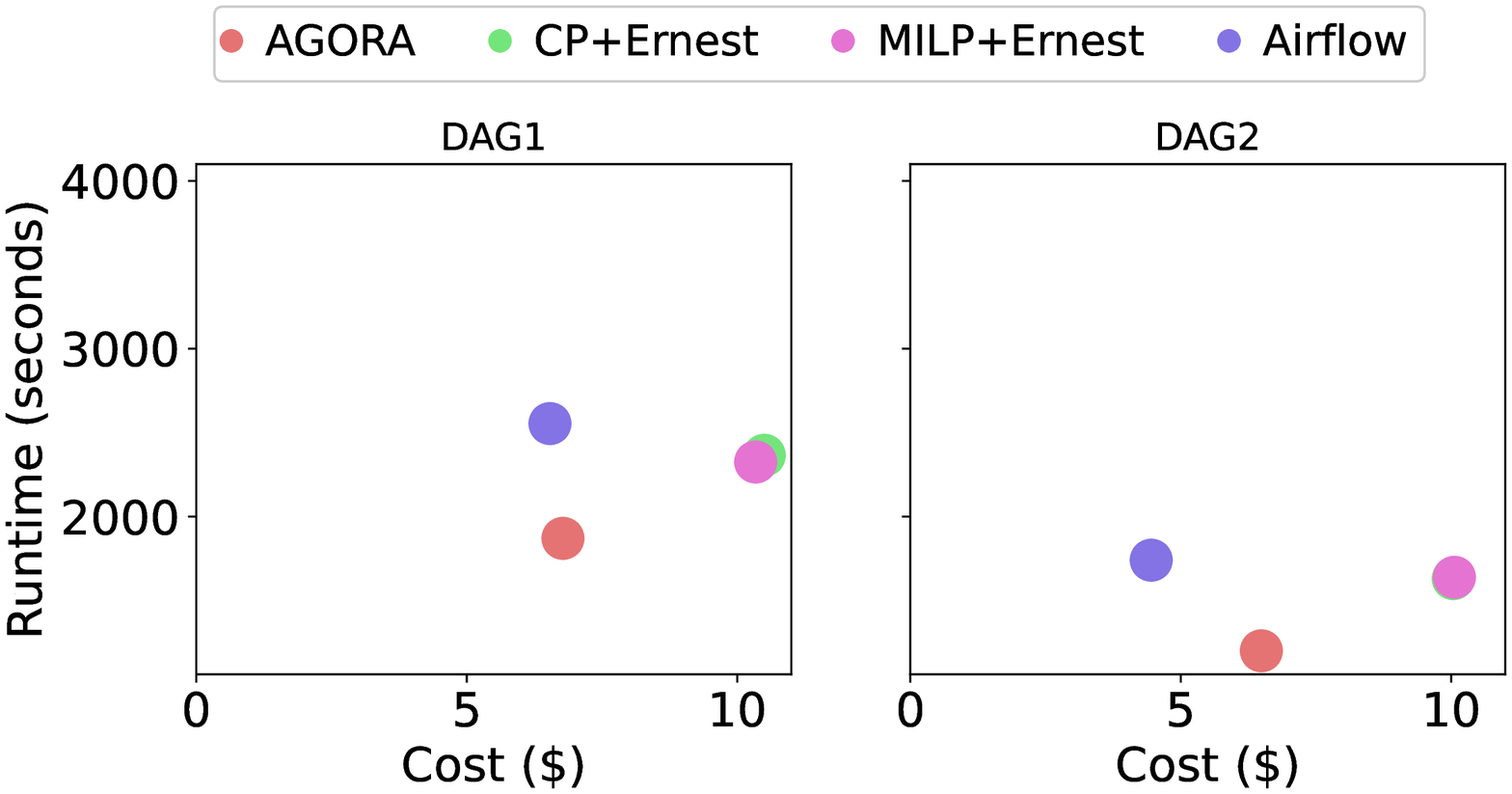}
    \caption{runtime}
    \end{subfigure}
       \begin{subfigure}[t]{0.32\linewidth}
    \centering
    \includegraphics[width=\linewidth]{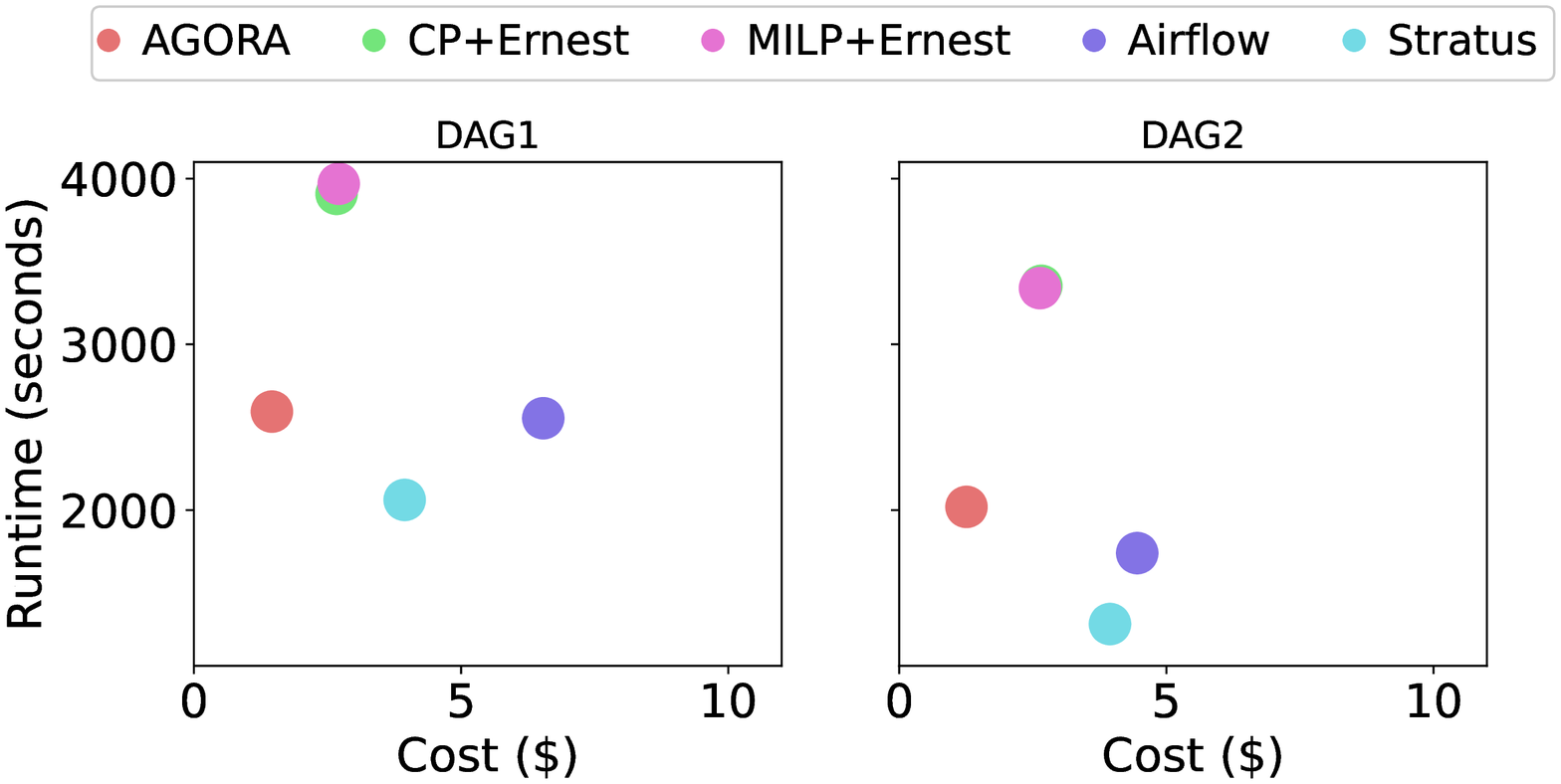}
    \caption{cost}
    \end{subfigure}
    \vspace{-8pt}
    \caption{End-to-end runtime and cost of studied DAGs under default Airflow, \proj, CP+Ernest, MILP+Ernest, and Stratus under different optimization goals: balanced, runtime, and cost. Y-axix shows the overall DAG runtime in seconds, and x-axis shows the cost. The lower left dots represent the better cost-performance.}
    \label{fig:eval-overall}
\end{figure*}
\begin{figure*}
    \begin{subfigure}[t]{0.5\textwidth}
    \centering
    \includegraphics[width=\linewidth]{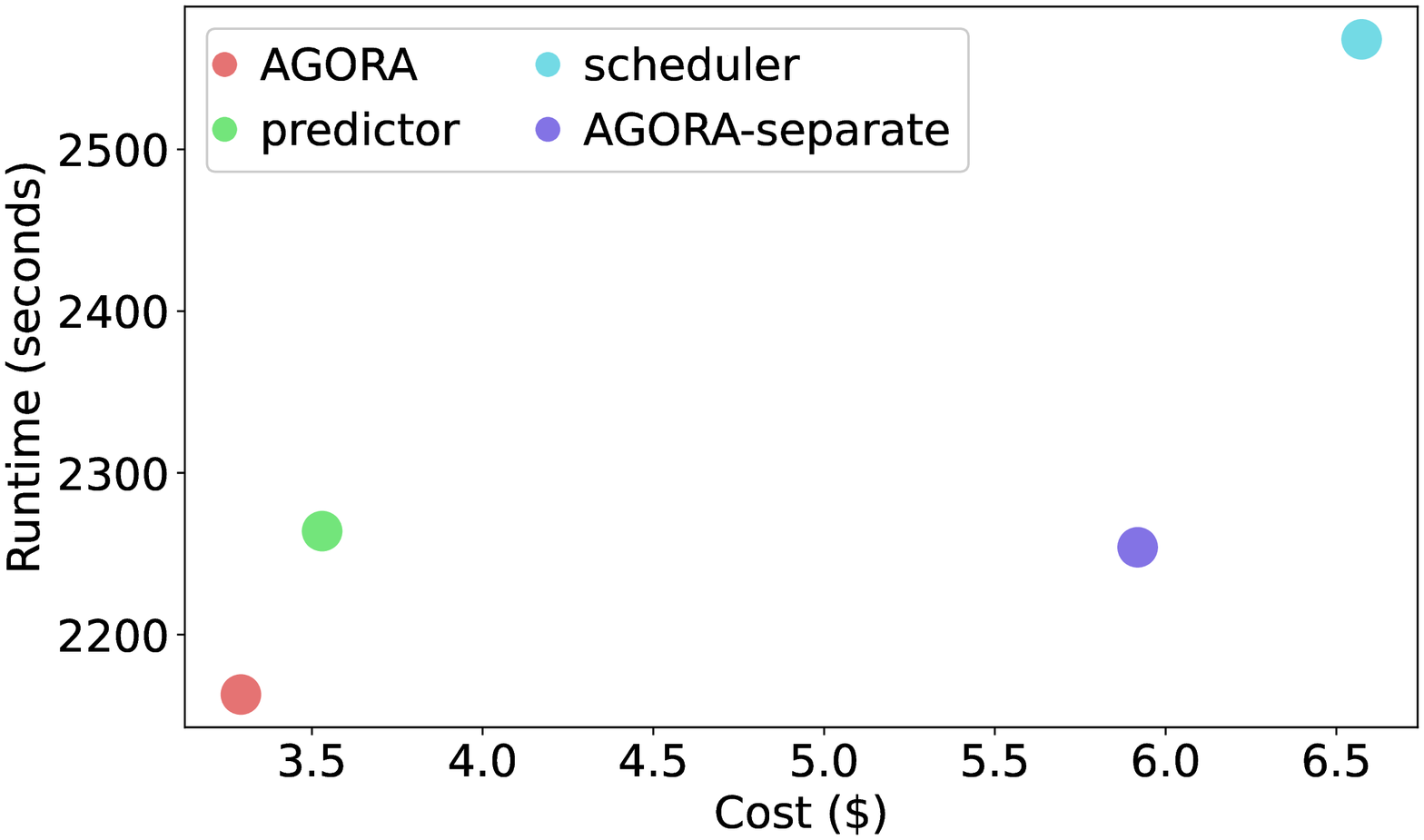}
    \caption{DAG1}
    \label{fig:exp3-dag1}
    \end{subfigure}
    \begin{subfigure}[t]{0.48\textwidth}
    \centering
    \includegraphics[width=\linewidth]{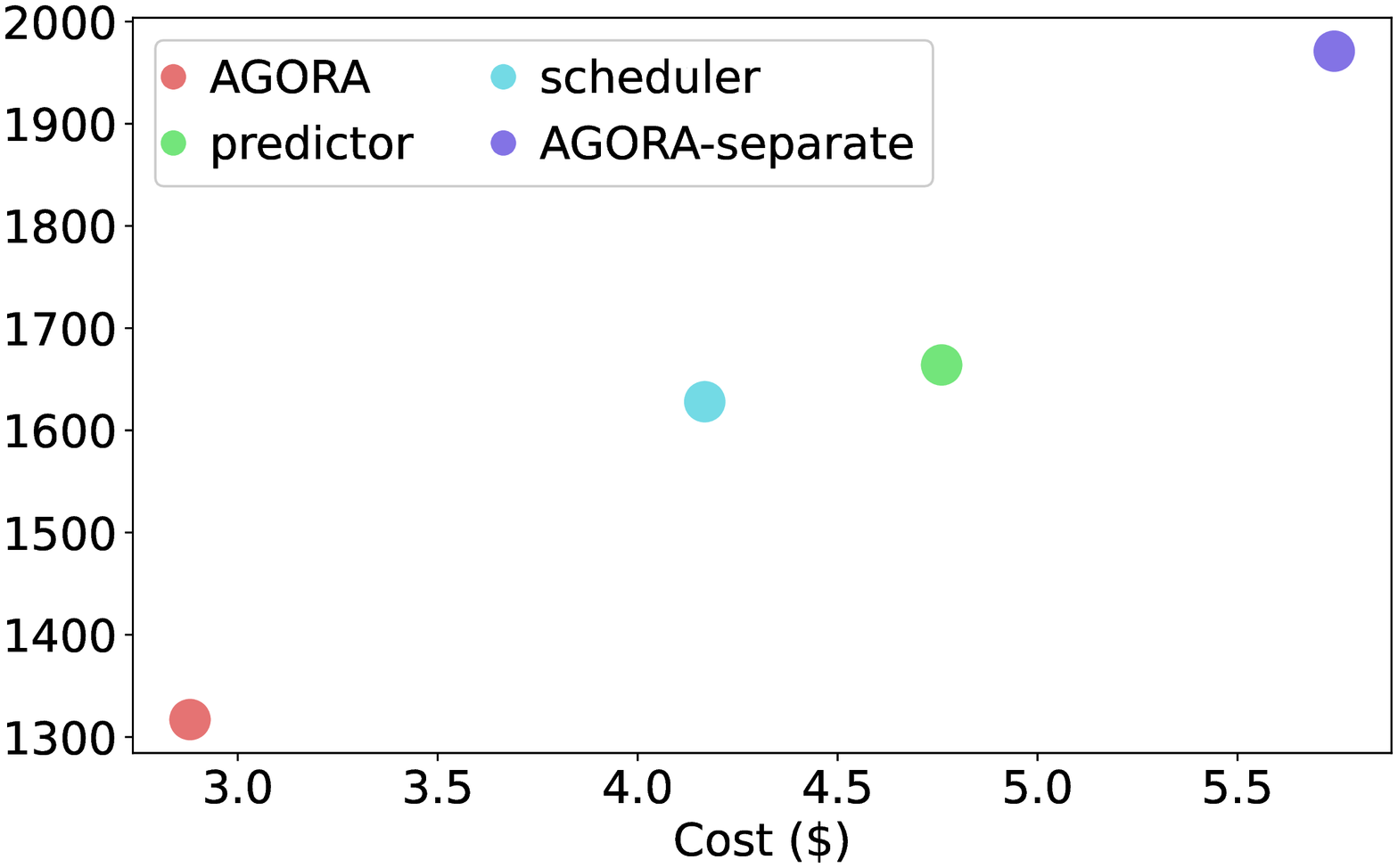}
    \caption{DAG2}
    \label{fig:exp3-dag2}
    \end{subfigure}
    \vspace{-10pt}
    \caption{Performance and cost of {\proj} with predictor only, scheduler only, and separately optimized (no co-optimization).}
    \label{fig:eval-breakdown}
    \vspace{-10pt}
\end{figure*}
In this experiment, we compare the overall performance and cost gains of {\proj} against other state-of-the-art solutions. Other than Apache Airflow, which has been widely adopted in industry, we have implemented Ernest~\cite{2016-ernest} for selecting the best VM configurations, together with Critical Path (CP~\cite{1969-cp}) as the representative of a heuristic based scheduler and Mixed Integer Linear Programming (MILP~\cite{1997-milp}) as the representative of an optimization based scheduler. We have implemented {\proj} and Ernest to optimize with three goals: balanced, runtime, and cost. For {\proj}, the runtime goal optimizes towards the lowest end-to-end runtime of the DAG; the cost goal optimizes for the lowest cost to run the DAG; and the balanced goal sets the weight ($w$) to $0.5$ to balance both runtime and cost. Note that {\proj} is also able to optimize for other goals by tuning the weight to shift between runtime and cost. On the other hand, Ernest can only optimize each task instead of the whole DAG. With this limitation, we have set Ernest to optimize each task with the three goals to pick the best runtime, cost, and a balanced runtime and cost. We also implemented Stratus~\cite{2018-stratus} specially for cost since Stratus is designed to only optimize for cost. We also embedded DAG dependencies into Stratus. While Airflow is not able to perform optimizations, it is widely used in real-world production environments. We add it here as a baseline for all goals as an anchor. We use default Airflow configurations which handle DAGs with topological order and a FIFO scheduler. We adopt the aforementioned Spark configurations (Section~\ref{sec:motivation}) for Airflow, Ernest, Stratus, and the initial input for {\proj}. 

Figure~\ref{fig:eval-overall} shows the results. We can see that {\proj} performs better than other methods across all three optimization goals. For the balanced goal, {\proj} achieves both better runtime and cost, with a runtime improvement of $15.3\%$ and $24.3\%$, and a cost improvement of $49.6\%$ and $35.3\%$ for DAG1 and DAG2 respectively. For the runtime goal, {\proj} optimizes for runtime only and yields a higher cost than the default Airflow by $3.5\%$ and $31.4\%$, but it improves the runtime by $36.6\%$ and $45.0\%$ for DAG1 and DAG2 respectively. For the cost goal, {\proj} has worse performance in terms of runtime compared with Ernest+CP and Ernest+MILP, but it results in the lowest cost with comparable runtime against the default Airflow by an improvement of $77.7\%$ and $71.7\%$ for DAG1 and DAG2 respectively. Although Stratus is designed to optimize cost, it still shows higher cost compared with {\proj} and Ernest. However, Stratus shows the lowest runtime in this experiment. This is because Stratus utilizes more resources eventually. Stratus does not attempt to determine the best tradeoff between cost and runtime, instead it simply utilizes any resources available and tries to minimize cost based on that. As a result, it eventually uses more resources than the other methods. In some cases (e.g., balanced goal for DAG2), we see that Ernest+CP and Ernest+MILP is worse than default Airflow which does not optimize. This is because separate optimization can only optimize individual tasks and cannot see the whole DAG instead of the global optimization of {\proj}, and it sometimes results in worse end-to-end performance than not optimizing at all. We see this phenomenon again later in Section~\ref{sec:breakdown}. In general, DAG1 takes longer time to run and costs more as well. We observe that DAG1 has a higher chance for cost improvement while DAG2 has more room for runtime improvement. This is because DAG1 has less parallelism than DAG2. For example, DAG1 has tasks that are waiting for a single task to finish before the other tasks begin (the top and second to last tasks in Figure~\ref{fig:eval_dag}), which is not present in DAG2. As a result, the single tasks might become a straggler for the whole DAG, but low parallelism also leads to less VM resource usage and lower costs. 

\subsection{Performance Breakdown}
\label{sec:breakdown}

In this experiment, we zoom in to examine the impact of each component in {\proj} (i.e., {\bf Predictor} and {\bf Scheduler}). For simplicity, we configure {\proj} to optimize for the balanced goal. We configure {\proj} to enable {\bf Predictor} only without the {\bf Scheduler} and vice-versa. We also change {\proj} to perform with both {\bf Predictor} and {\bf Scheduler} optimized separately (without co-optimization) similar to solutions using Ernest in previous experiments (referred to as {\proj}-separate). Results are shown in Figure~\ref{fig:eval-breakdown}. We see that for DAG1, the {\bf Predictor} contributes more than the {\bf Scheduler}, but the opposite for DAG2. This is same as our observation in the first experiment (Section~\ref{sec:overall}) that DAG2 has a more complicated DAG structure and higher parallelism for the scheduler to yield a bigger contribution. Meanwhile DAG1 has tasks dependent on single task which will benefit more from appropriate resource allocation performed by the {\bf Predictor}. However, naively enabling both without co-optimization ({\proj}-separate in the figure) does not result in the best performance. For example, {\proj}-separate is worse than {\bf Predictor} only for DAG1, and it shows the worst performance for DAG2. This is again consistent with our observation in Section~\ref{sec:overall} that in some cases, it is worse to optimize separately. We can see from the figure that compared to the case without co-optimization, {\proj} (with co-optimization) is $4.0\%$ faster and $44.4\%$ cheaper for DAG1, and $33.8\%$ faster and $49.8\%$ cheaper for DAG2. This underscores our idea of co-optimization in this paper.  

\subsection{Optimization Goals}
\begin{figure}[t]
\centering
\includegraphics[width=\linewidth]{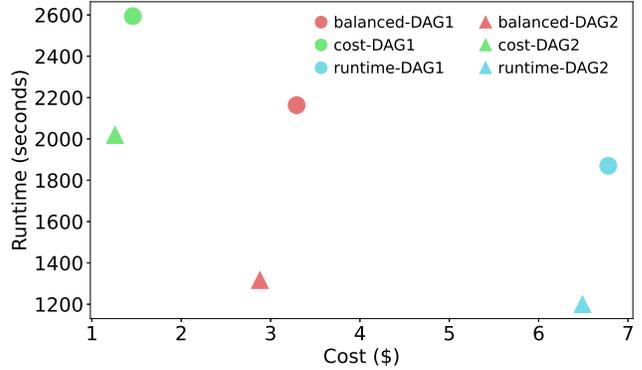}
\vspace{-20pt}
\caption{Cost and performance of {\proj} under different optimization goals. Circle dots represent DAG1, and triangle dots represent DAG2.}
\label{fig:eval-goals}
\end{figure}
In our next experiment, we further dig into the capability of {\proj} to optimize towards different goals. Note that {\proj} is able to fine-tune the optimization focus with a sliding weight $w$ described in Section~\ref{sec:design-problem}. For simplicity, we only select three common goals: balanced ($w=0.5$), cost ($w=0$) and runtime ($w=1$). We see in Figure~\ref{fig:eval-goals} that for both DAG1 and DAG2, cost goal points reside in top left corner which is cheapest but has the highest runtime. In contrast, the runtime goal points in the lower right corner demonstrate the highest cost with lowest runtime. Finally, the balanced goal looks for a middle point in the figure with a balanced tradeoff between the two. From this figure we can also see that different DAGs show a different optimization trend. Here we see that DAG2 demonstrates a stiffer curve than DAG1, confirming our observation that DAG2 has more opportunity for runtime optimization. In reality, many DAG workflows are even more complex and have highly variable characteristics. These results underscore the ability of {\proj} to fine-tune optimization preferences.

\subsection{Optimization Overhead}
\label{sec:eval-overhead}
\begin{figure}[t]
\centering
\includegraphics[width=\linewidth]{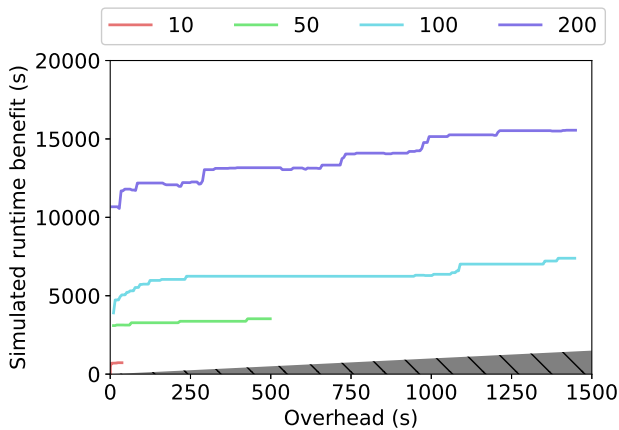}
\vspace{-20pt}
\caption{Trade-off between overhead and runtime benefits for different number of tasks and DAGs. Shaded area indicates when overhead $\geq$ runtime benefits.}
\label{fig:overhead_sim}
\end{figure}

With optimization of any NP-hard problem, one key concern is always the overhead associated with the optimization. Often, the cost of performing the calculation outweighs any potential benefits. In this experiment, we evaluate the optimization overhead of {\proj} with a large-scale simulation. We simulate with randomly generated DAGs with a width of 4 and a depth of 3-5 consisting of 10 tasks each. We increase the number of DAGs from 1 to 20 to result in 10 to 200 total tasks. Figure~\ref{fig:overhead_sim} shows the trade-off between optimization overhead and predicted runtime improvements. We can see from the figure that the overhead associated with increasing problem sizes increases from tens of seconds to over a thousand seconds, but the runtime benefit also increases dramatically as well from hundreds of seconds to over $15,000$ seconds, showing that for all problem sizes the optimization benefits are worth the overhead time. For small problems ($\leq 50$ tasks), the solution converges quickly. For larger problems, the optimization can be stopped earlier to get most of the runtime benefits with less overhead. The shaded area in the figure indicates the region when the overhead is greater than or equal to runtime benefits. Clearly, none of the problem sizes is in that region. 

In all of our previous experiments, we observe optimization overheads of around $35$ seconds for DAG1 and $45$ seconds for DAG2, which only counts for less than $2\%$ of total runtime of the DAG. In addition, {\proj} is designed to allow for offline optimization which can eliminate the optimization overhead during runtime.

Although optimization overhead is negligible, there is still opportunity for {\proj} to speed up the solver. While in this paper, the solver has a serial implementation on a single core, the algorithm is friendly to parallel computing. In our future work, we plan to explore the parallelization of our solver, and possibly adopt emerging hardware accelerators to further speed up the algorithm. The SAT formulation of the scheduling problem allows emerging specialized hardware systems to be designed and implemented that could dramatically reduce the solve time~\cite{chou_analog_2019, molnar_accelerating_2020}. With the increasing complexity of data pipelines, we see the inevitable need to speed up {\proj} and a broad range of emerging technologies capable of doing so.

\subsection{Macro-Benchmark}
We have been experimenting with {\proj} using micro-benchmarks and demonstrating promising results. However, we are also interested in how {\proj} scales and performs in real-world production environments. To this end, we use a real-world large-scale production trace from Alibaba~\cite{2020-alibaba} and apply {\proj} with simulation. 
\subsubsection{Methodology}
Alibaba's cluster handles two types of jobs: online services and batch workloads. Since {\proj} is designed to improve scheduling of DAG workloads, we focus on the batch DAG workloads in the trace. More specifically, the 2018 Alibaba cluster trace is used in this simulation. The 2018 Alibaba cluster trace includes jobs run on 4034 machines over a period of 8 days. 
There are over 4 million jobs (represented as DAGs) and over 14 million tasks. Each machine has 96 cores and an undisclosed amount of memory. Memory requests for each task are given in percentages of a machine's memory. A previous analysis of the cluster trace showed that $95\%$ of the online services utilize less than $20\%$ of the CPU cores, and less than $40\%$ of memory of cluster~\cite{guo_who_2019}. In our simulation, we reasonably reduce the cluster size by $20\%$ for CPU and $40\%$ for memory as the usage for the DAG jobs. 

For each task, the trace provides the required cores, required memory, and task runtimes. Using this data, we generate a random scaling curve for each task using the universal scalability law (USL~\cite{gunther_hadoop_2015}). The universal scalability law is a model that accounts for the concurrency, contention, and coherency of the system, extending Amdahl's law with the coherency parameter which accounts for delays from crosstalk. We chose USL because it is general and accounts for negative scaling behavior that can occur in some distributed compute jobs. We use the generalized form of the USL with three parameters, randomly choosing $\alpha$ and $\beta$ for each task, and calculating $\gamma$ to fit the demands requested and runtime data provided by the trace:
\begin{equation}
    X(N) = \frac {\gamma N}{1 + \alpha(N-1) + \beta N(N-1)},
\end{equation}
where $X$ is the throughput, $N$ is the number of cores, $\gamma$ is the concurrency parameter, $\alpha$ is the contention parameter, and $\beta$ is the coherency parameter. Each parameter is bound between 0 and 1. 

The cluster trace also provides submission times of each job. In this simulation, {\proj} is triggered to schedule jobs that have been submitted every fifteen minutes or when the demands in the queue are greater than three times the available cores in the cluster. Therefore, {\proj} can respond to ad-hoc DAG submissions by users and does not require the same set of DAGs to always be submitted together. This also demonstrates the capability of {\proj} to support multi-DAG optimization.

The total runtime to finish all DAG workloads in the trace and the total cost associated are the two metrics in this simulation.

\subsubsection{Result}
\begin{figure}[t]
\centering
\includegraphics[width=\linewidth]{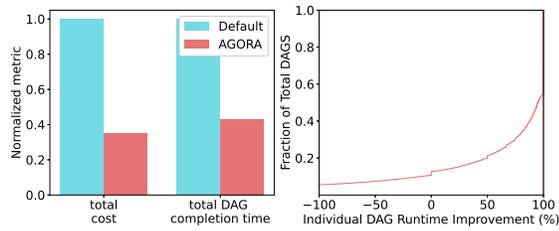}
\caption{Normalized Cost and total DAG completion time (left) and the CDF of runtime improvements for each individual DAG (right).}
\label{fig:alibaba_sim}
\end{figure}

Figure~\ref{fig:alibaba_sim} shows the results. {\proj} reduces the total cost of the workload by $65\%$ and the total completion time by $57\%$. Looking at the completion time of each DAG separately, $87\%$ of the DAGs have runtime improvements, with $45\%$ of the DAGs having nearly $100\%$ improvement. Therefore, the users who submitted the DAGs mostly see a benefit in their DAG completion time, and the organization (if running these workloads in a public cloud environment) also sees an overall benefit in cost of the workloads.

\section{Conclusion}
This paper introduces the concept of of co-optimizing resource configuration and job scheduling for running DAG workflows in cloud environments. We also design {\proj}, an automated globally optimized resource allocator and scheduler to demonstrate significant cost and performance gains of co-optimization while maintaining negligible overheads. The unique ability to fine-tune optimization preferences allows {\proj} to optimize according to the diverse characteristics of workloads and the particular needs of the users. {\proj} simplifies running data pipelines in the cloud by relieving users of the burden of choosing resource configurations from a sea of choices, and could eventually create a seamless and serverless-like experience. 

\bibliographystyle{ACM-Reference-Format}
\bibliography{references}

\end{document}